# Analysis methods and code for very high-precision mass measurements of unstable isotopes

J. Karthein[1,2,*,@], D. Atanasov[1], K. Blaum[2], D. Lunney[3], V. Manea[3], M. Mougeot[1]

1: *CERN, 1211 Geneva 23, Switzerland*
2: *Max-Planck-Institut für Kernphysik, 69117 Heidelberg, Germany*
3: *IJCLab, CNRS Université Paris-Saclay, 91406 Orsay, France*

*: present address: *Massachusetts Institute of Technology, MA-02139 Cambridge, USA*
@: corresponding address: karthein@mit.edu

**ABSTRACT** — We present a robust analysis code developed in the Python language and incorporating libraries of the ROOT data analysis framework for the state-of-the-art mass spectrometry method called phase-imaging ion-cyclotron-resonance (PI-ICR). A step-by-step description of the dataset construction and analysis algorithm is given. The code features a new phase-determination approach that offers up to 10 times smaller statistical uncertainties. This improvement in statistical uncertainty is confirmed using extensive Monte-Carlo simulations and allows for very high-precision studies of exotic nuclear masses to test, among others, the standard model of particle physics.

**INTRODUCTION** — One of the fundamental observables of the atomic nucleus is the nuclear binding energy. This quantity directly results from the difference of the nuclear mass and the sum of the individual nuclear constituent masses (neutrons and protons). While measurements of stable systems allow for long preparation and measurement times, the precision for short-lived nuclides is strongly limited by the production rate at the radioactive ion beam (RIB) facilities (down to a few atoms per hour), as well as the half-lives (down to a few tens of milliseconds). However, studying these exotic systems often provides exclusive insights into the nuclear structure, the weak interaction or rapid nucleosynthesis processes compared to stable atoms. This particularity motivates technical advances in the field of high-precision mass spectrometry. Over the past four decades, Penning traps have emerged as the experimental tool of choice for high-precision measurements of atomic masses [Bl06, Hu17]. The ISOLTRAP spectrometer has pioneered the technique for radioactive species [Kl13] and Penning trap facilities now exist at all radioactive ion beam facilities worldwide [Lu18]. Further developments in the measurement technique using ISOLTRAP [Bl02, Ge07] allowed us to recently set a new world record for the highest precision on a mass of a short-lived ($T_{1/2}$ < 1 h) isotope [Ka19a]. But this experiment also demonstrated the technique's intrinsic precision limitation, requiring hundreds of ions per measurement cycle, which is not feasible for the most short-lived isotopes of interest for nuclear physics.

Groundbreaking new measurement technique for Penning traps was developed by S. Eliseev *et al.* [El13] at GSI based on preliminary studies by Eitel *et al.* [Ei09] in Mainz. This so-called phase-imaging ion-cyclotron-resonance (PI-ICR) technique allows for the direct mass determination in a non-scanning approach. It thus allows for up to 25 times faster measurements [El14, Ka19b] at the same or better levels of precision.

This technique was successfully implemented at the ISOLTRAP setup [Ka19b, MaKa20] for which an analysis code was developed in Python [Ka19c]. Given an optimal beam characteristics and high statistics, a new analysis approach presented below allows for a multifold reduction of the statistical uncertainty compared to the state of the art without any intervention on the experimental apparatus. It can even be applied to all already-existing datasets and opens the door for very-high-precision mass determination required for stringent tests of the Standard Model with uncertainties in the order of a few (tens) eV/$c^2$.

**PI-ICR THEORY** — The mass determination in a Penning trap is based on the direct determination of the cyclotron frequency $\nu_c$,

$$\nu_c = \nu_+ + \nu_- = \frac{B}{2\pi} \times \frac{q}{m} \,.$$

In the case of an ideal Penning trap, $\nu_c$ is also equal to the sum of the frequencies of the radial eigenmotions, $\nu_+$ and $\nu_-$, of the trapped ion [Br86] (also, see Fig. 3). The magnetic field strength $B$ can be eliminated with the ratio $r$,

$$r = \frac{\nu_{c,\mathrm{ref}}}{\nu_{c,\mathrm{ioi}}} \,,$$

between the cyclotron frequency $\nu_{c,\mathrm{ref}}$ of a well-known reference ion and the cyclotron frequency $\nu_{c,\mathrm{ioi}}$ the ion of interest as long as these consecutive measurements are performed within a time-frame shorter than non-linear fluctuations of the magnetic field (typically ~15 min) [Ke03].

In order to reduce the statistical uncertainty, many of such measurement pairs are performed under different measurement conditions to eliminate eventual systematic effects. The final ratio $r$ is calculated by fitting simultaneously the same polynomial function $f_n(t)$ describing the temporal evolution of the cyclotron-frequency measurements, $\nu_{c,\mathrm{ioi}}$ and $\nu_{c,\mathrm{ref}}$, respectively:

$$\nu_{c,\mathrm{ioi}}(t) = f_n(t) \,,$$
$$\nu_{c,\mathrm{ref}}(t) = r \times f_n(t) \,. \qquad (1)$$

The degree $n$ of the polynomial function $f_n(t)$ is adjusted to the lowest degree describing the environment-temperature-induced fluctuations of the magnetic field. One can minimize the reduced $\chi^2_{\mathrm{red}}$ of the data distribution to the fit, which typically results in the same polynomial degree $n$. This ratio-determination method has proven to be very robust [Na93, Ka19a, Ka19b, Fi12, El15] while minimizing over-fitting of the model is therefore crucial. In this work, a second-order linear non-orthogonal ion-optical transfer polynomial is used to describe the response of model parameters such as the total travelling time of the ion to the detector. This modelling is depicted in Fig. 4 for an example event.

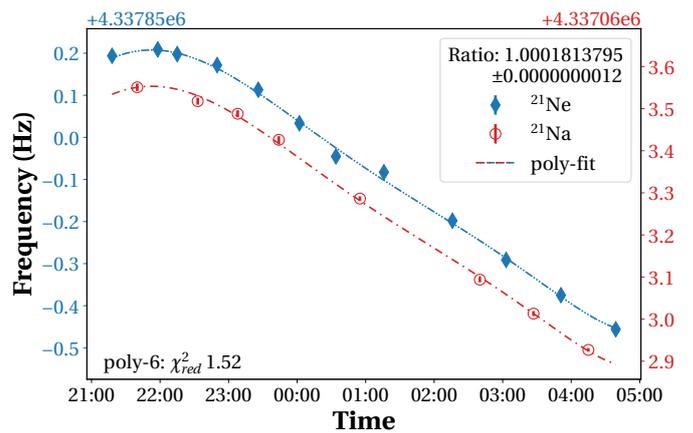

Fig. 1: Frequency-ratio determination based on the temporal cyclotron-frequency distributions of the ion of interest (red) and the reference ion (blue), as well as simultaneous polynomial fits. The frequency scale differs between the two species to allow for a better display of both data sets. The displayed data is a subset of the data published in [Ka19a]. For details, see text.

The atomic mass $m_{ioi}$ of the ion of interest

$$m_{ioi} = r \times (m_{ref} - m_e - E_{B,ref}/c^2) + m_e + E_{B,ioi}/c^2, \quad (2)$$

can finally be determined by comparing the cyclotron frequencies of the two ionic species in the Penning trap, correcting for the reference ion mass $m_{ref}$, the mass of the electron $m_e$ as well as the respective electron binding energies $E_B$ after the ionisation of either atom. While $m_e$ and $E_B$ of light elements are typically known to high precision, they can no longer be neglected at the precision level achieved in this work.

The PI-ICR measurement technique makes use of the fact that the mass determination can be reduced to a radial frequency determination in a Penning trap. In PI-ICR, a radial frequency $\nu_r$ is derived from the full phase $\varphi_{full}$ accumulated by an ion in a very precisely measured time $t_r$:

$$\nu_r = \frac{\varphi_{full}}{2\pi t_r} = \frac{2\pi n + \varphi_r}{2\pi t_r}.$$

This full phase consists of an integer number $n$ of full turns $n \in \mathbb{N}_0$ as well as an additional phase $\varphi_r \in [0, 2\pi[$ between the starting point and the end point of the motion. In practical situations, the integer number of turns can be determined based on a prior (much less precise) knowledge of the ion's mass, but this knowledge can also be built up entirely by PI-ICR. This can be achieved by starting with the time $t_r$ as a fraction of one full period $T_r = (t_n - t_0)/n$ and then stepwise increasing $t_r$. The more $t_r$ is increased, the smaller also the uncertainty on $\nu_r$ becomes since the statistical uncertainty of $\nu_r$ in this case only originates from the determination of the additional phase $\varphi_r$. One full period for the radial eigenmo-tions in a typical Penning-trap setup is on the order of:

$$T_{\nu_+} \sim \mu s \ ; \ T_{\nu_-} \sim ms \ .$$

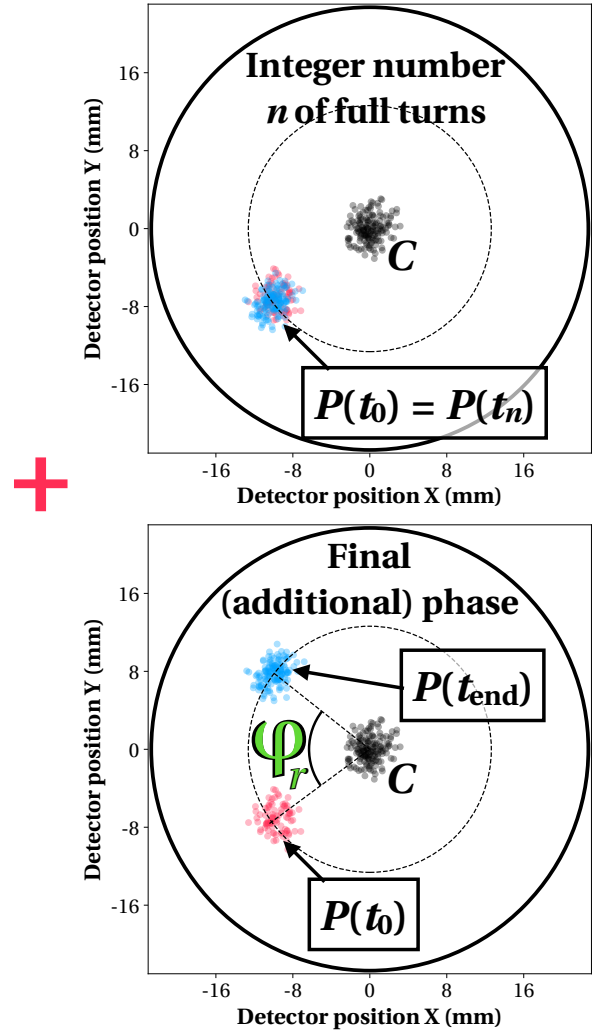

Fig. 2: Graphical description of the determination of the full phase $\varphi_{full}$ for a radial motion using the PI-ICR technique. The full phase of a trapped ion consists of an integer number of full turns $n \in \mathbb{N}_0$ (top) as well as a final, additional phase $\varphi_r \in [0, 2\pi[$ (bottom). This final phase $\varphi_r$ of the radial motion extends between the spot position $P(t_0)$ (red) of the ions at the starting time $t_0$, the position of the ions $P(t_{end})$ (highlighted in blue) after the evolution time $t_r = t_{end} - t_0$, and a reference point $C$ representing the center (black) of the radial motion (dashed line).

This basic procedure is depicted in Fig. 2 where the additional phase $\varphi_r$ (highlighted in green) is defined by the position $P(t_0)$ (highlighted in red) of the ions at the starting time $t_0$, the position of the ions $P(t_\text{end})$ (highlighted in blue) after the evolution time $t_r = t_\text{end} - t_0$ inside the trap at frequency $\nu_r$, as well as a reference point $C$ (highlighted in black) representing the center of the radial motion (dashed line). In case of a Penning trap, a measurement scheme was developed by S. Eliseev *et al.* [El14] for the direct determination of the cyclotron frequency $\nu_c$ in only three measurement steps:

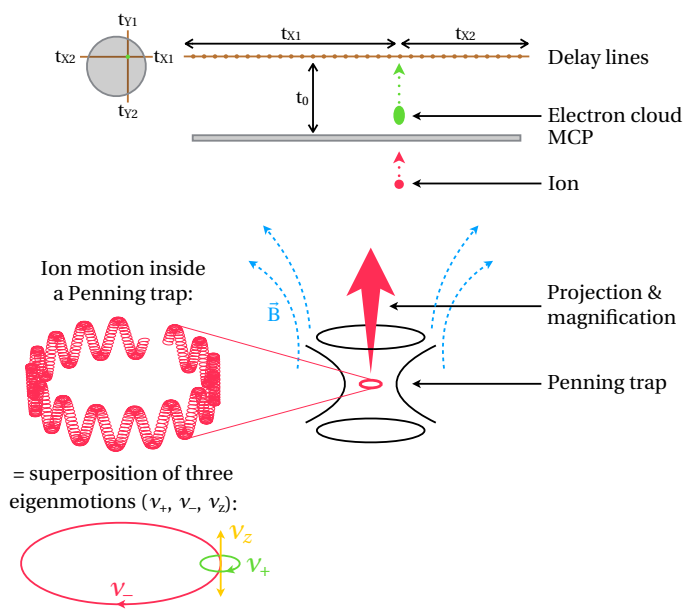

Fig. 3: PI-ICR detection principle. The ion motion inside the Penning trap is projected onto the detector while being magnified due to the magnetic field gradient. At the detector, the individual ion is converted into an electron cloud when hitting the micro-channel plate (MCP). The resulting signal triggers the first time event. The electron cloud is then accelerated onto a delay-line array (here: Roentdek DLD40), where the induced signal in the X and Y wires propagates towards their extremities, generating in total four additional time stamps. These five time stamps allow for a precise position detection.

1. The ion motion is cooled by radial centering in the Penning trap via resonant RF-excitation [Kr99]. This position is captured by ejecting the ion onto an on-axis 2D-position-sensitive micro-channel-plate (MCP) detector, which is parallel to the radial eigenmotions of the ion inside the trap (see Fig. 3). This step allows for a calibration of the radial motions and only has to be repeated once every couple of hours since this position is very stable over time [El14].

2. The ion is prepared on a pure $\nu_+$ motion by resonant RF-excitation [Kr99] and after evolution at $\nu_+$ for precisely the accumulation time $t_{acc}$ the position is recorded as in step 1. Different ions can be prepared over several cycles at exactly the same starting point $P_0(x,y)$ in order to be able to repeat this step for increased statistics and thus reduced statistical error. Before ejection onto the detector, the ion motion is converted into a (slow) $\nu_-$ motion by RF-excitation to reduce the angular spread on the detector. This conversion hereby conserves the absolute value of the accumulated phase but switches the sign of the direction of the phase evolution [El14], hence the plus sign for $\varphi_\text{final} = \varphi_+ + \varphi_-$ in Eq. (3).

3. The ion is prepared on a pure $\nu_+$ motion as in step 2 and immediately converted into a pure $\nu_-$ motion by resonant RF-excitation [Bl03, Ei09]. After evolution at $\nu_-$ for precisely the same accumulation time $t_{acc}$ the position is recorded as in step 1 and 2. It is important to also use the same starting point $P_0(x,y)$ as for step 2 in order for Eq. (3) to be valid.

Each of these three measurement steps is repeated multiple times per measurement cycle in order to allow for a determination of the mean position of the so-called spots on the detector (see Fig. 2). These steps reduce the determination of the cyclotron frequency $\nu_c$,

$$\nu_c = \nu_+ + \nu_- = \frac{2\pi n_+ + \varphi_+}{2\pi t_{\text{acc}}} + \frac{2\pi n_- + \varphi_-}{2\pi t_{\text{acc}}}$$

$$= \frac{2\pi (n_+ + n_-) + \varphi_{\text{final}}}{2\pi t_{\text{acc}}}, \qquad (3)$$

to the determination of the final phase $\varphi_{\text{final}}$, defined by the spots in steps 1, 2 and 3 instead of requiring one additional spot for each frequency at $t_0$ thanks to the same preparation scheme.

***ANALYSIS CODE — RECONSTRUCTION*** The analysis is divided into three parts. First, the raw data is read and transformed into position information. Second, the position information is used to determine the total phase and the cyclotron frequency. Third, the mass of the ion of interest is calculated based on the temporal distribution of cyclotron frequencies.

Thanks to the Jupyter notebook environment [Kl16], all parts of the code are easy to read while being fairly customizable. They are independent and can be adjusted to the individual requirements of different experiments or used for a different measurement technique. As an example, the third part concerning the determination of the mass based on the simultaneous fitting technique can also be used for all Penning-trap mass spectrometry techniques based on frequency-ratio determination, such as the time-of-flight ion-cyclotron-resonance (ToF-ICR) technique in it's single-pulse [Gr80] or Ramsey-type application [Ge07] with no adjustments.

The first part of the code deals with the event-position reconstruction by importing the five time stamps (MCP, X1, X2, Y1, Y2) from the LabView data acquisition. These timestamps are visualized in Fig. 3. The MCP time is used for a calibration of the four time signals in the X and Y direction of the detector by a simple subtraction. Next, the difference between the two time stamps per direction gives the position on the detector along this direction. The detector should be calibrated once by stepwise increasing the excitation amplitude of one of the radial ion motions until no ions are detected. This allows to set the proper "time windows" for the individual time stamps marking the edges of the detector.

Since it is possible to capture more than one ion inside the trap, the data acquisition at ISOLTRAP is configured such that all events in one measurement sequence are saved in their individual channel (MCP, X1, X2, Y1, Y2; see Fig. 3) after being converted to NIM standard by a constant-fraction discriminator.

Due to the pulse-height distribution of the signals and limits on minimum discriminator threshold, sometimes one or two of the five required time stamps per position are missing. This leads to a disturbance in the order of incoming events, which is why a computationally demanding comparison between all individual time stamps per measurement sequence has to be performed. This calculation ensures finding the real events within all acquired time stamps by performing the sum check described above.

The combinatory computation is performed with a cascaded for-loop, which compares each time stamp in all of the five channels with each other. Smart if-conditions are imposed to skip false events and thus reduce the calculation time. This part of the code is CPU-parallelized where appropriate. A hardware-accelerated (GPU) alternative using 5-dimensional meshgrid tensors (one dimension each for X1, X2, Y1, Y2, MCP, and thus replacing the need for the combinatorics based on for-loops), tensor masking (replacing the comparison- and if-conditions) and just-in-time compilation, all based on the machine learning library JAX from Google [Br18], was tested and found to be about 10 times slower on the available office-PC hardware due to the much-increased number of computations per event. But since the data acquisition is limited in measurement time to ensure linearity in the magnetic-field drifts (on the order of ~15 min), the position reconstruction typically only takes about one second per file for about tens or hundreds of files in a typical experiment. The position information is then saved in a new *.csv-file such that this step only has to be performed once.

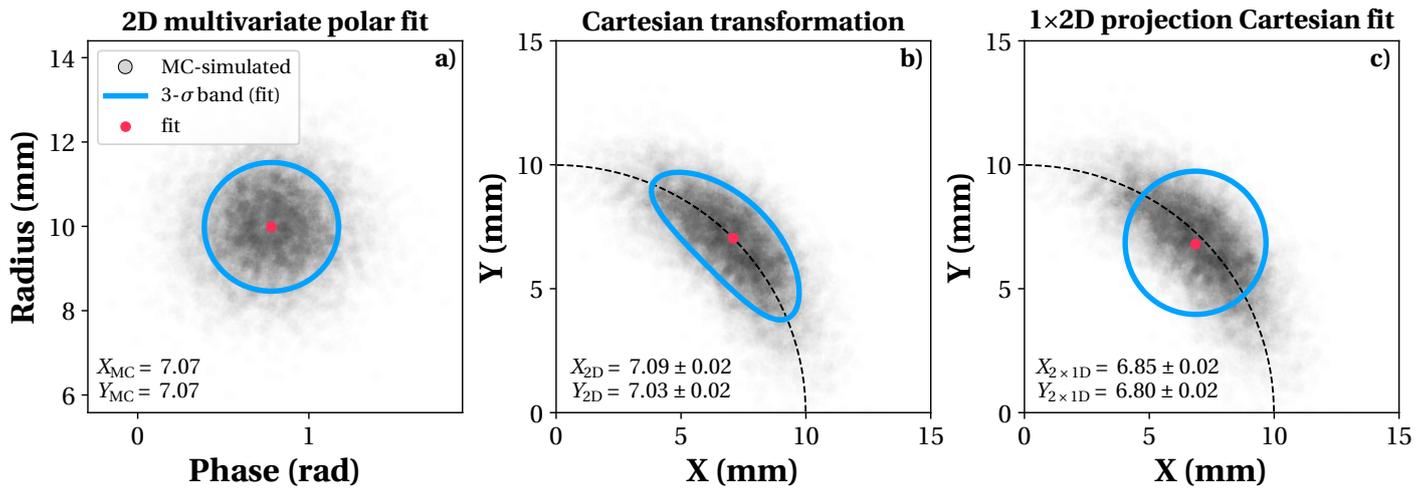

Fig. 4: Multivariate Gaussian 2D-fit for a Monte-Carlo (MC) simulated circular distribution in polar coordinates (a) and its transformation into Cartesian coordinates (b). For a visual comparison, the corresponding 2×1D Gaussian fit is shown in c). Its deviation from the theoretical values $X_{\mathrm{MC}}, Y_{\mathrm{MC}}$ compared to the fit uncertainties (without correction for correlation in case of the projection fit in c) is significant while the multivariate fit in polar coordinates (a, b) agrees with the theoretical values $X_{\mathrm{MC}}, Y_{\mathrm{MC}}$. Spot shapes like the one shown in b) can occur, e.g., due to timing jitters during the ejection procedure or in the direct projection of the (fast) $\nu_+$ motion.

**ANALYSIS CODE — FREQUENCY EVALUATION** — The next part in the analysis is to translate the position information into a phase information and from that to calculate the cyclotron frequency of the ions. It was purposely chosen to perform the analysis on a file-by-file basis rather than a batch analysis to ensure full control over all cuts and fits. Furthermore, the Pandas library [Mc10] was used extensively to ensure maximal readability and adjustability of the code. In a first step, several analysis cuts can be performed:

- *ToF-cut*: The time-of-flight (ToF) distribution from the trap to the detector allows for a limitation in the mass of interest to remove contaminants of different masses. ToF outliers are reduced automatically within a 10-Sigma range.

- *Position cut*: The position region of interest can be limited to the desired area of the detector (e.g., to cover only the blue spot in the top part of Fig. 2). Moreover, position outliers are detected and reduced automatically within a 10-Sigma range of the mean position. This cut also allows for the removal of contaminating species (such as single isomeric states; see Ref. [MaKa20]).

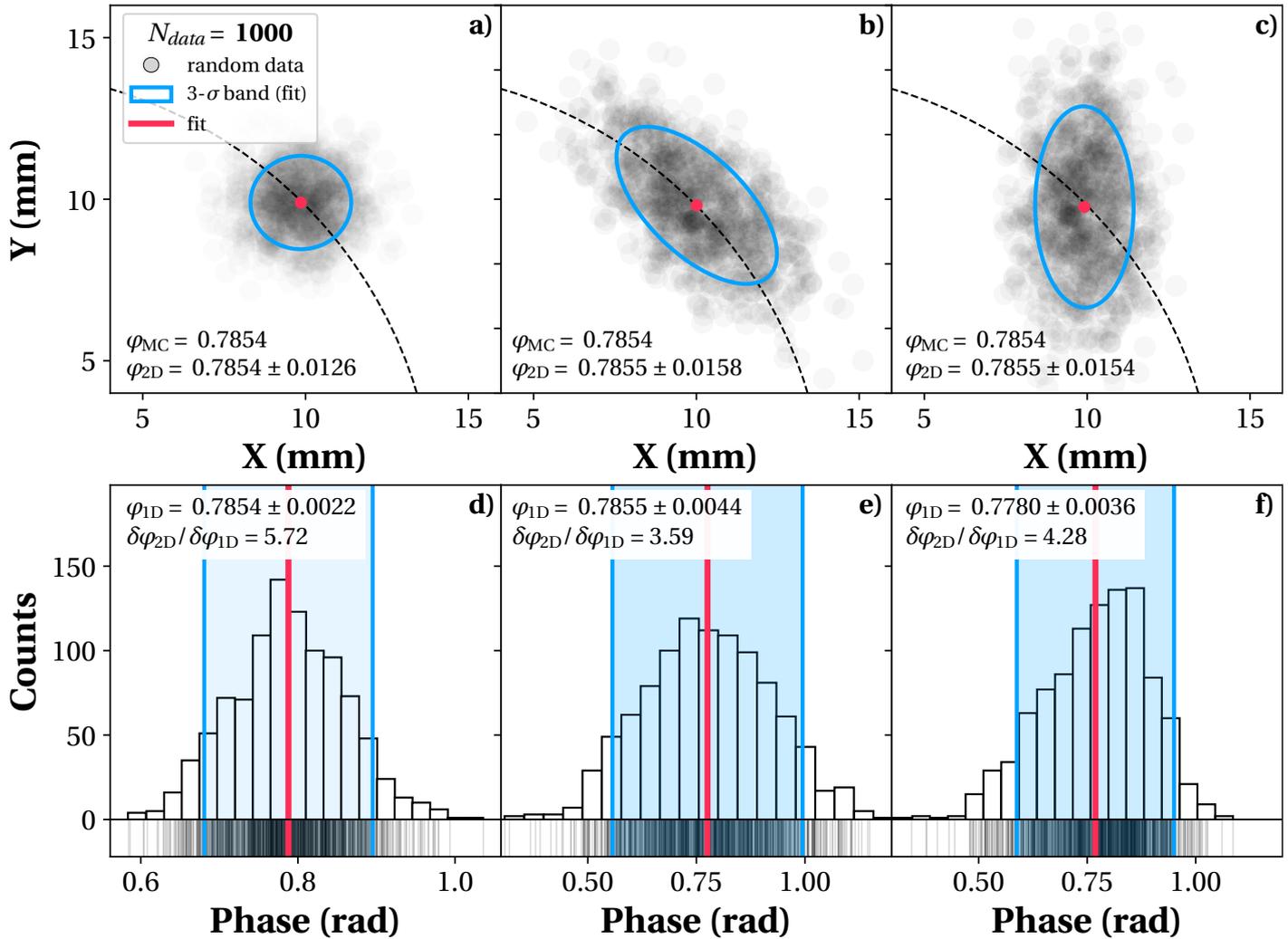

Fig. 5: Monte-Carlo (MC) simulation of three-spot scenarios for a symmetric, aligned elliptical, and rotated elliptical (with respect to the ion motion along the dashed circle) 2D Gaussian distribution with 1000 ions, as well as their multivariate fit in 2D (top row) and their phase-projected 1D fit (bottom row) in comparison. Elliptical spots as shown in a) and b) can occur due to asymmetries and imperfections in the trap or transport potentials. The true phase $\varphi_{MC}$ is compared to the average phase $\varphi_{2D}$ resulting from the mean X/Y coordinates from the multivariate 2D fit and the average phase $\varphi_{1D}$ from the 1D fit. Both averages are calculated from the fit results of 1000 random MC samples for each of the three distributions. The given uncertainties correspond to the average one-sigma uncertainty for the given quantity over the 1000 simulations. Additionally, the reduction factor in statistical uncertainty from the 2D to the 1D fit is shown for the three spot scenarios. The symmetric and aligned spots agree well with the true phase value, whereas the 1D fit for the rotated elliptical spot shows a >2 sigma deviation from the true value.

- *Z-cut*: Limiting the number of ions per capture to reduce space-charge effects [Ke03].

In each step, the number of removed spots are printed transparently into the output console.

Next, the positions for the reference, $\nu_+$, and $\nu_-$ spots have to be determined. The approach so far with the SHIPTRAP LabView code [Ka20] was using a 2×1D Gaussian model in X and Y directions to fit the unbinned position data (see Fig. 4c). In our approach, we are using a 2D multivariate Gaussian model with the maximum likelihood estimation for the unbinned position data. The difference between the two models is that the multivariate version properly accounts for two-dimensionality of the distribution with its rotation in space, whereas the regular 2×1D Gaussian model merely fits the individual projections onto each dimension. In a perfect experiment with a circular distribution in the X-Y-space there will be no dif-ference between the two models. Deviations will however occur in reality. If we assume an exaggerated case with a highly elliptical or asymmetric spot at a rotation of 45 degrees (as depicted in Fig. 4b), a 2×1D Gaussian model would result in a circular distribution instead of an elliptical one since it essentially fits only the projections in X and Y. The phase of both fits might agree in case of a symmetric distribution, but the 2D-projection Gaussian model naturally results in highly correlated fitting parameters, which increase the off-diagonal elements of the fit's covariance matrix. These correlations lead mathematically to a significant additional contribution in the statistical uncertainty, which typically is not accounted for.

In the multivariate case on the other hand, correlation for PI-ICR spots at any ellipsoidal deformation or rotation at any point of the circle is negligible due to the intrinsic rotation matrix in the model. Hence, the spot in Fig. 5b can be properly described as well as any other spot in the simulation. This model was chosen over the proposed alternative approaches described below due to its robustness and versatility for any spot shape. It builds on the spatial distribution based on initial buffer-gas cooling, as derived in Eq. (12) *ff.* in [El14].

The minimization of the negative log-likelihood function for the maximum likelihood estimation is performed with the iMinuit library [On12], which ports the Minuit2 minimizer of CERN's analysis framework ROOT [Br97] to Python. It was chosen over different SciPy [Vi20] minimizers due to its robustness and direct support at CERN. The library allows in addition the integration of automatic differentiation, i.e., using exact derivatives instead of numerical derivation in the minimization as available, e.g., in the JAX library [Br18]. This integration is planned in a future performance update, but is not essential for the analysis code at this stage.

From the positions of the reference, $\nu_+$, and $\nu_-$ spots one can determine the phase $\varphi_{\text{final}}$ with the arctangent function "atan2" of the NumPy library [Ha20a], which has the whole circle as image set instead of just $]-\pi/2, \pi/2[$. This phase finally allows for the calculation of the cyclotron frequency $\nu_c$ according to Eq. (3). The number of full turns $n_{+,-}$ for each eigenfrequency has to be predetermined as described above using the stepwise increase of the accumulation time $t_{acc}$.

All analysis data from the raw events to the fitting information is saved in a pickle-file (*.p) for traceability and transparency. All results are additionally saved in a comma-separated-value file (*.csv) for the independent calculation of the frequency ratio and then the mass.

***ANALYSIS CODE — FREQUENCY-RATIO*** — The calculation of the final mass is performed based on the exact time stamp of the measurement and the calculated $\nu_c$ and can thus also be used for different Penning-trap mass spectrometry techniques like the ToF-ICR technique in its single-pulse or Ramsey-type excitation scheme since they also use the calculated $\nu_c$ and the exact time stamp of the measurement for the frequency-ratio determination. The software uses a polynomial function $f_n(t)$ according to Eq. (1) to describe the temporal distribution of all measured $\nu_c$-values using the least squares method (see Fig. 1). The statistical uncertainty is calculated from the covariance matrix of the fit, which is performed for all possible polynomial degrees $n \in [1, \#_{\nu_c} - 1]$ with the number of $\nu_c$ data points $\#_{\nu_c}$ to avoid overdetermining the fit. The software automatically chooses the fit with a $\chi^2_{\text{red}}$ closest to one for the final ratio $r$, but the degree could also be fixed, e.g., to the known degree (if known) of the temporal fluctuations during the time of the measurement. The results of this analysis step are saved in a separate *.csv-file. From the final ratio $r$, the mass $m_{\text{ioi}}$ can be calculated according to Eq. (2), using the latest literature mass for the reference ion $m_{\text{ref}}$ [Hu17]. It is common to publish the final ratio $r$ to update $m_{\text{ioi}}$ if ever a more precise $m_{\text{ref}}$ is available.

**ALTERNATIVE APPROACH 1: STRONG RADIAL DEFORMATION** — In case of strong radial deformation of the Gaussian distribution in Cartesian coordinates along the ion motion (as depicted in Fig. 4c), a transformation for all spot positions into polar coordinates is highly recommended. As emphasized by Fig. 4a, the same distribution can be described much better in this coordinate system. Applying a regular multivariate Gaussian model to the data in Cartesian coordinates would result in a deviation to the actual mean position of the spot. In the code, a simple transformation to polar coordinates can be performed at the beginning of the "Spot fitting" section as well as back to Cartesian coordinates after the fitting. By this, no further adjustments have to be made to the code. An example is given in the subfolder "alternative-approaches".

**ALTERNATIVE APPROACH 2: VERY HIGH-PRECISION** — In the following, we present an alternative analysis approach that allows — given the proposed preparation — up to 10 times smaller statistical uncertainty on the same dataset compared to the robust multivariate approach described above. This is a game changer for many applications, such as Weak-interaction Standard Model tests based on the determination of the $V_{\text{ud}}$ element [Ka19a]. Given the exact same experimental operation and high statistics, relative statistical uncertainties of $\delta\nu_c/\nu_c \leq 10^{-10}$ can be reached for the first time in a few hundred milliseconds measurement time only.

The analysis approach is based on a transformation of the spot positions into polar coordinates. This transformation results in a position information separated into a phase coordinate and a radial coordinate instead of X and Y. Since in PI-ICR one is interested in the mean position of the distribution along the phase coordinate, the approach only uses the phase information and neglects the radial position information. This reduction has the advantage that the number of free parameters in the maximum likelihood estimation reduces from five, in case of a multivariate 2D Gaussian model (mean X, mean Y, sigma X, sigma Y, rotation angle used in the covariance matrix) to only two free parameters (mean phase, sigma phase).

We performed extensive Monte Carlo (MC) simulations to study the effect and accuracy of this new approach for many different, realistic scenarios by varying the following properties:

- spot shapes: ellipticity from 1 (Fig. 5a) to 2 (Fig. 5b/c); larger ellipticities are not recommended due to lack of control over systematics
- rotation with respect to the tangent to the circular trajectory of the spot center: 0 (see Fig. 5a/b) to 45 degrees (see Fig. 5c); behavior repeats afterwards
- position on the detector: 0 to 45 degrees with respect to the horizontal axis and counted counterclockwise; behavior repeats afterwards

- ratio between trajectory radius and spot diameter: 10 mm / 2 mm up to 15 mm / 2 mm
- number of ions per spot: 10, 100, 1000, 10000

Each MC simulation was repeated 1000 times in order to calculate average fit result values, average fit uncertainties, the variance of the individual fit results, and the skewness of the fit results scattering to ensure the statistical significance of the following statements. An example code is given in the subfolder "alternative-approaches".

The results clearly demonstrate the power of this approach as long as the spot is reflection symmetric with respect to a radial axis passing through the trajectory center. This is depicted in Fig. 5a/d and 5b/e. The top row shows the initial unbinned MC-random distribution (here for 1000 ions per spot) and their unbinned maximum likelihood estimation using the multivariate 2D Gaussian models described in the "ANALYSIS CODE — FREQUENCY EVALUATION"-section. Their results are displayed in red for the mean X/Y position and in blue for the 3-sigma ellipse. The evaluation agrees with the true value while providing a small statistical uncertainty for the positions parameter. From the average X and Y position information, a phase can be calculated as described in the same section mentioned above.

In the bottom row, the position information displayed in the top row is transformed point by point into polar coordinates and the unbinned phase information is shown. In case of experimental data, the position information would first need to be calibrated by the position of the center of the circular motion. This average position of the measurement's step 1 must be determined in any approach first with the 2D multivariate fit. The phase information is also displayed in a binned manner to guide the eye. The results of the unbinned maximum likelihood estimation of the 1D phase data is displayed in red for the mean phase position and in blue for the 3-sigma range. For the mirror-symmetric cases with respect to the radial axis, i.e., column one and two, both the new approach and the multivariate approach agree on average in the phase determination with the theoretical value for any number of simulated ions per spot: 10, 100, 1000, 10000. However, the new approach reaches much-reduced statistical uncertainties as displayed in the uncertainty ratio $\delta\varphi_{2D}/\delta\varphi_{1D}$ between the 2D approach versus the 1D approach. These reduction factors for the average statistical uncertainty found in the MC simulations performed for all mentioned spot conditions are presented in Tab. 1 for different numbers of ions per spot.

The simulations were performed for 10000, 1000, 100 and 10 ions to cover the range of statistics found in a realistic experiment. For a very low number of ions (10-100) as would be accumulated in a typical experiment at the production limitation of present RIB facilities [Ka20] (few ions per second/minute/hour), the reduction in statistical uncertainty is moderate between a 20% and over 300% improvement. However, it is much more interesting for experiments at higher production rates where typical Standard Model tests are performed. In these situations where much more data can be accumulated, significant improvement factors on the statistical uncertainty of 4-10 can be achieved just by applying the proposed fitting approach.

Tab. 1: Statistical uncertainty improvement factors for the 1D-phase fitting approach compared to the 2D multivariate Gaussian fitting approach for different number of ions and circular/elliptical mirror-symmetrical shapes (see left and middle columns of Fig. 5). The improvement factors are calculated for the mean statistical uncertainties of 1000 MC simulations. The improvement factor ranges correspond to the evaluation at different angles (0-45 degrees) and realistic ratios between the trajectory radius and spot diameter (2σ) of 5-8.

| $\#_{ion}$ | Improvement factors | |
|---|---|---|
| | circle | ellipse |
| 10 | 1.8 | 1.2 - 1.3 |
| 100 | 3.2 | 2.0 - 2.3 |
| 1000 | 5.6 - 5.7 | 3.6 - 4.0 |
| 10000 | 9.8 - 10.1 | 6.4 - 7.1 |

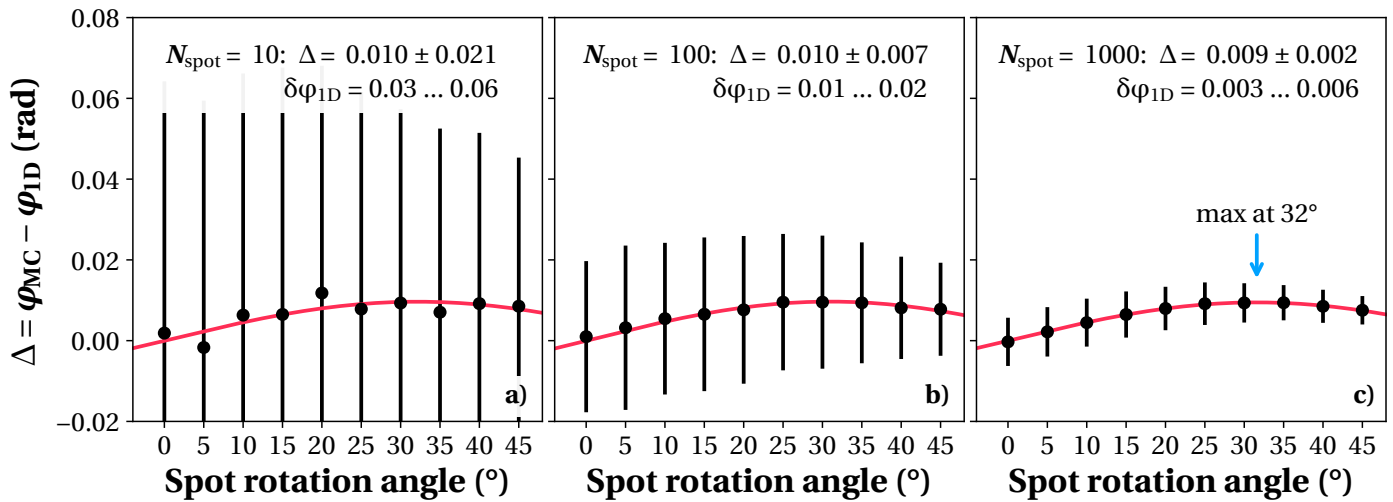

Fig. 6: Systematic absolute deviation of the 1D phase fit for an elliptical spot of ellipticity 2 and for different spot rotation angles with respect to the the tangent to the circular trajectory of the spot center. The results are shown for 10 (a), 100 (b) and 1000 (c) ions per spot. Each deviation represents the mean of 1000 MC simulations and is plotted with its standard deviation as error bar. The behavior is compared to a sine function (red) from which the maximal deviation $\overline{\Delta}$ (~0.1 rad) and the rotation angle for the maximal deviation (~32 degrees) were derived. For comparison, the mean individual statistical uncertainty $\delta\varphi_{1D}$ for 1000 repeated MC simulations is shown with its range for the given spot rotation angles.

While the robust 2D multivariate fitting approach can be applied for any type of spot deformation introduced by different kinds of systematic effects (e.g., jitter in the timing, fringe fields, or edge effects in the transport to the detector, fluctuations in the potentials, ...), the 1D phase fitting approach will result in a systematic deviation from the true value as soon as the spot is no longer symmetric to the motion axis. This situation is emphasized in Fig. 5c/f for a spot with 45 degree rotation and an ellipticity of 2 (in case of stronger deformed spots, better ion preparation as described in Ref. [El14] is recommended).

This systematic absolute deviation is shown in Fig. 6 for different spot rotation angles and different number of ions per spot. Each data point represents the mean deviation for 1000 MC simulations and their standard deviation as error bar. Independent of the number of ions in the spot, a maximal absolute deviation of ~0.01 rad was found for the most extreme deformation and rotation. The resulting relative shift for the cyclotron frequency of an ion of mass m ~ 100 u in a Penning trap of 6 T (→ $\nu_c$ ~1 MHz) measured at typical 100 ms and 1000 ms phase accumulation time would amount to $1.6 \times 10^{-8}$ and $1.6 \times 10^{-9}$, respectively. This is on the order of the latest estimate on the upper limit of the systematic uncertainty of $8 \times 10^{-9}$ for the ISOLTRAP setup as described in Ref. [Ke03]. This limit will be revisited during planned systematic studies in the following months. Thus, in typical mass determinations of short-lived isotopes with low production yields (few ions per second/minute/hour) and short half-lives (~100s ms), the moderate statistical improvement factors of the 1D phase approach between 20% and over 300% might be at the cost of a possible additional systematic uncertainty from the fit which is on the same order of magnitude. Since these few number of ions per spot make a symmetry judgement difficult even via skewness calculations of the distribution, the robust 2D multivariate approach is recommended.

A different conclusion can however be made for so-called *Q*-value measurements, in which not the actual mass but the mass difference between two isobars is determined. These measurements are typically used for high-precision Standard Model tests since they allow for significant cancellation of most systematic effects. Hereby, the two ions of interest of same mass number are consecutively prepared at the exact same spot in the trap und thus undergo the exact same systematic shifts. While high statistics would make an optimization for symmetry easy (and is in any situation highly recommended), the fit-induced systematic shift in the individual mass determination would however cancel out even for asymmetric and rotated spots since the shift will be on the same order and direction for both isobars (rigorously true at least for spot distortions arising from transport optics effects). In summary, this new phase fitting approach presents a clear advantage over other existing and the presented robust 2D phase determination method specifically for *Q*-value measurements. This is true for any number of ions and any spot shape with statistical uncertainties up to 5-10 times smaller than with existing techniques.

***CONCLUSION/OUTLOOK*** — The recently developed PI-ICR method has become the state-of-the-art mass spectrometry method for radioactive isotopes. In this publication, two new analysis schemes were presented for the first time, offering up to 10 times smaller statistical uncertainties even on existing datasets. This improvement opens the window for next-generation Standard Model tests and significant contributions to the determination of the neutrino mass via the end point in the electron-capture spectrum using short-lived isotopes at low energy.

***ACKNOWLEDGEMENT*** — We acknowledge financial support by the Max-Planck Society. JK acknowledges financial support by a Wolfgang Gentner Ph.D. Scholarship of the BMBF (05E15CHA) and the ISOLDE collaboration for transitional support during the COVID-19-pandemic.